\documentclass[aps,prl,onecolumn]{revtex4}
\usepackage{amssymb,epsf}
\usepackage{latexsym}
\usepackage{graphicx}
\usepackage{epsf}
\usepackage{epstopdf}
\usepackage{epsfig}
\usepackage{hyperref}
\usepackage{color}

\begin{document}

\title{Near-extremal black holes in Weyl gravity:\\
Quasinormal modes and geodesic instability}
\author{ Mehrab Momennia$^{1}$\footnote{
email address: m.momennia@shirazu.ac.ir} and Seyed Hossein Hendi$^{1,2}$%
\footnote{%
email address: hendi@shirazu.ac.ir}} \affiliation{$^1$ Physics
Department and Biruni Observatory, College of Sciences, Shiraz
University, Shiraz 71454, Iran\\
$^2$ Research Institute for Astronomy and Astrophysics of Maragha (RIAAM),
P.O. Box 55134-441, Maragha, Iran}

\begin{abstract}
Stability criteria of nearly extreme black holes in the perturbations level
is one of the interesting issues in gravitational systems. Considering the
nearly extreme conformal-de Sitter black holes, in this paper, we obtain an
exact relation for the quasinormal modes of scalar perturbations. As a
stability criteria, we find a lower bound on the event horizon radius which
is corresponding to an upper bound on the value of the quasinormal
frequencies. In addition, we show that the asymptotic behavior of
quasinormal modes gives highly damped modes which is important due to
possible connection between their real part and the Barbero-Immirzi
parameter. We also obtain the Lyapunov exponent and the angular velocity of
unstable circular null geodesics. Finally, we examine the validity of the
relation between calculated quasinormal modes and unstable circular null
geodesics.
\end{abstract}

\maketitle

\section{Introduction}

The detection of gravitational waves (GWs) of black hole binary mergers \cite%
{Abbott1,Abbott2,Abbott3} by the LIGO and VIRGO observatories opened a new
window to the multimessenger astronomy. Besides, more recently, the Event
Horizon Telescope collaboration released the first image of the `shadow' of
a supermassive black hole in the galaxy M87 \cite{Akiyama1,Akiyama4}. This
image alongside the emitted GWs proof the existence of black holes in the
cosmos which is one of the most important achievements of general relativity.%
\textbf{\ }Perhaps more remarkably, investigation of more realistic black
hole solutions helps ones to obtain other interesting aspects of black holes
based on a more successful theory than Einstein gravity.

One of the most interesting and successful theories in higher derivative
gravity scenario is so-called conformal gravity (Weyl gravity) defined by
the square of the Weyl tensor \cite{Bach,Buchdahl,Riegert,Mannheim,Pope}.
Among diverse modifications of general relativity, Conformal Gravity (CG)
theory is special because it enjoys a Weyl invariant action. It is also
unique up to the choice of the matter source and coupling constants in order
to keep the Weyl invariance properties. In addition, CG is a
higher-curvature theory of general relativity which is power-counting
renormalizable \cite{Stelle,Faria}, and therefore, one can consider it as a
suitable theory for constructing quantum gravity \cite{Bergshoeff,Wit}.
Although this theory of gravity suffers the Weyl ghost, it is possible to
remove it under certain conditions \cite%
{MannheimDavidson,Bender,BenderMannheim,BenderMannheim2008,BenderMannheimPRL,Mannheim2013,MannheimPLB,Mannheim2018}%
.

It is worth mentioning that CG can be examined as a possible UV completion
of Einstein's general relativity \cite{Adler,Hooft,MannheimFound}. It also
arises from twister-string theory \cite{Witten}, and as a counterterm in
adS/CFT calculations \cite{Tseytlin,Henningson,Balasubramanian}. In contrast
with the Schwarzschild solution, we can apply the CG to explain the rotation
curve of galaxies without considering the dark matter \cite%
{MannheimRC,BrienRC,MannheimBrienRC}, and therefore, one may hope to
streamline the conception of dark side of the universe \cite{MannheimFound}.
It is notable that $4-$dimensional solution of Einstein gravity is also a
solution of CG, and there is an equivalence between Einstein gravity and CG
by considering the Neumann boundary conditions \cite{Maldacena,Anastasiou}.

In order to have a stable black hole, one should investigate its behavior
under dynamic and thermodynamic perturbations to obtain the stability
criteria. The instability conditions are sufficiently strong to veto some
models. The quasinormal modes (QNMs) are the resulting behavior of black
holes whenever they undergo dynamic perturbations \cite%
{Chandrasekhar,Kokkotas,Berti,Konoplya}. The QNMs describe the evolution of
fields on the background spacetime and the QNM spectrum of gravitational
perturbations can be observed by gravitational wave detectors. Therefore,
investigation of QNMs attracted much attention during the past three years
after the detection of the gravitational radiation of compact binary mergers
by LIGO and VIRGO observatories \cite{Abbott1,Abbott2,Abbott3}.

The idea of QNMs was started first in $1957$ with the work of Regge and
Wheeler to investigate the stability of black holes under small
perturbations \cite{Regge}. The frequency of perturbations has been
calculated by using a semi-analytic approach \cite{Schutz,Iyer,KonoplyaWKB}
and several numerical methods \cite%
{ChandrasekharDetweiler,Leaver,Horowitz,Naylor}. Studying the perturbations
of black hole spacetime bring a lot of interest in a large area of physics:
thermodynamic properties of black holes in loop quantum gravity \cite%
{Dreyer,Kunstatter,Motl}, AdS/CFT correspondence \cite%
{Horowitz,CardosoLemos,Birmingham,KonoplyaAds,Starinets,MossNorman},
dynamical stability of compact objects \cite%
{Gundlach,HodPRD,Cardoso,Delsate,Brito,Cuyubamba,KonoplyaZhidenko2017,Stuchlik}%
, the possible connection with critical collapse \cite%
{Horowitz,KonoplyaPLB,Oh} and the null geodesics \cite%
{Press,Goebel,Ferrari,Mashhoon,Nollert,BertiKokkotas,Geodesic,KonoplyaGeo}.
In case of conformal gravity, the astrophysical gravitational waves of
compact binaries has been investigated in \cite{Caprini,Holscher}. Here, we
are going to study the scalar perturbations of nearly extreme conformal-de
Sitter (conformal-dS) black holes.

The quasinormal frequencies (QNFs) corresponds to the QNMs are independent
of the initial perturbations and describe the black hole response to the
perturbations on the background of spacetime. The QNFs are related to the
black hole charges, such as mass and electric charge, and therefore, they
reflect the properties of background geometry. On the other hand,
investigation of the geodesic motions of test particles and null geodesics
around black holes are an interesting astrophysical phenomena since the high
curvature properties of black holes have considerable effect on the geodesic
motions. Perhaps, one may think of a possible relation between the QNFs and
geodesics around a black hole. Cardoso and his colleagues claimed that the
QNMs of any stationary, spherically symmetric, and asymptotically flat black
hole in the eikonal limit can be determined by the parameters of the
circular null geodesics \cite{Geodesic}. In this regard, the real part of
QNMs is a multiple of the circular null geodesic frequency $\Omega _{c}$ and
their imaginary part is a multiple of the inverse of the instability
timescale (Lyapunov exponent $\lambda $) associated with this geodesic
motion. The correspondence between the QNFs in the eikonal limit ($%
l\rightarrow \infty $) and circular null geodesics is given by
\begin{equation}
\omega =\Omega _{c}l-\left\vert \lambda \right\vert \left( n+\frac{1}{2}%
\right) i;\ \ n=0,1,2,...,  \label{correspondence}
\end{equation}%
where $l$ and $n$ are, respectively, the multipole number and the overtone
number. Although this correspondence works for a number of cases \cite%
{Geodesic,KonoplyaGeo}, it may violate \cite{Stuchlik} whenever the
perturbations are gravitational type or the test fields are non-minimally
coupled to gravity \cite{KonoplyaGeo}. It has been shown that such
correspondence does not hold for Gauss-Bonnet gravity (which is a
higher-curvature modification of general relativity) for arbitrary coupling
constant \cite{KonoplyaGeo}.

In this paper, we shall study another higher-curvature modified gravity,
i.e. conformal (Weyl) gravity. We also obtain an analytical expression for
the QNMs of the conformal-dS black holes in the near-extremal regime by
considering a massless scalar perturbation minimally coupled to gravity.
Then, we check the validity of the correspondence between the QNMs and
circular null geodesics for this black hole. Since the scalar perturbation
is minimally coupled to a $higher-curvature~type$ of gravity, investigating
the correspondence (\ref{correspondence}) is a nontrivial task. In addition,
the conformal black holes in asymptotically flat spacetime reduce to the
Schwarzschild solutions which have been investigated in Refs. \cite%
{Geodesic,SchwDS}.

\section{review of the four-dimensional conformal solutions}

Here, we give a brief review of the four-dimensional black hole solutions in
Weyl gravity. The action of Weyl gravity is given by \cite{PangPope}
\begin{equation}
I=\frac{1}{16\pi }\int d^{4}x\sqrt{-g}C^{\mu \nu \rho \sigma }C_{\mu \nu
\rho \sigma },  \label{I3}
\end{equation}%
where $C_{\mu \nu \rho \sigma }$ is the Weyl conformal tensor. The field
equations are obtained by taking variations with respect to the metric $%
g_{\mu \nu }$%
\begin{equation}
\left( \nabla ^{\mu }\nabla ^{\nu }+\frac{1}{2}R^{\mu \nu }\right) C_{\rho
\mu \nu \sigma }=0,  \label{GFE}
\end{equation}

It is straightforward to show that the following $4$-dimensional line
element satisfies the field equations (\ref{GFE})
\begin{equation}
ds^{2}=-f(r)dt^{2}+f^{-1}(r)dr^{2}+r^{2}\left( d\theta ^{2}+\sin ^{2}\theta
d\varphi ^{2}\right) ,  \label{Metric}
\end{equation}%
where the metric function is as follows
\begin{equation}
f\left( r\right) =C_{1}+\frac{C_{2}}{r}+\frac{C_{1}^{2}-1}{3C_{2}}%
r+C_{3}r^{2},  \label{MF}
\end{equation}%
in which $C_{1}$, $C_{2}$ and $C_{3}$ are three integration constants. Here,
we redefine the constants of the metric function as $C_{2}=-2M$ and $%
C_{3}=-\Lambda /3$ for future comparisons with the Schwarzschild-dS black
hole. Based on the mentioned redefinition, the metric function takes the
following form
\begin{equation}
f\left( r\right) =C_{1}-\frac{2M}{r}-\frac{C_{1}^{2}-1}{6M}r-\frac{\Lambda }{%
3}r^{2}.  \label{MF1}
\end{equation}

It is notable that in contrast with the Einstein gravity, in which the
cosmological constant should be considered in the action by hand, it is
appeared as an integration constant in the conformal gravity. It is
worthwhile to mention that one can recover the Schwarzschild-dS solution by
setting $C_{1}$ equals one.

Here, we consider a massless scalar perturbation in the background of the
black hole spacetime and in the coming section, we will obtain an exact
formula for QNFs in the nearly extreme regime. The equation of motion for a
minimally coupled scalar field is given by
\begin{equation}
\square \Phi =0.  \label{SP}
\end{equation}

It is convenient to expand the scalar field eigenfunction $\Phi $ in the
form
\begin{equation}
\Phi \left( t,r,\theta ,\varphi \right) =\sum_{lm}\frac{1}{r}\Psi _{l}\left(
r\right) Y_{lm}\left( \theta ,\varphi \right) e^{-i\omega t},  \label{EXP}
\end{equation}%
where $Y_{lm}\left( \theta ,\varphi \right) $ denotes the spherical
harmonics and $e^{-i\omega t}$ shows the time evolution of the field.
Substituting the scalar field decomposition (\ref{EXP}) into (\ref{SP})
leads to a wavelike equation for the radial part $\Psi _{l}\left( r\right) $
as the following form
\begin{equation}
\left[ \partial _{r_{\ast }}^{2}+\omega ^{2}-V_{l}\left( r_{\ast }\right) %
\right] \Psi _{l}\left( r_{\ast }\right) =0.  \label{WE}
\end{equation}

In this equation, $V_{l}\left( r_{\ast }\right) $ is the effective potential
and $r_{\ast }$ is the known tortoise coordinate which are given by
\begin{equation}
V_{l}\left( r_{\ast }\right) =f\left( r\right) \left[ \frac{l\left(
l+1\right) }{r^{2}}+\frac{f^{\prime }\left( r\right) }{r}\right] ,
\label{EP}
\end{equation}%
and
\begin{equation}
r_{\ast }=\int \frac{dr}{f(r)},  \label{TC}
\end{equation}%
where $l$ and $\omega $ are, respectively, the multipole number and the
frequency of perturbations. We shall analyze the characteristic QN spectra $%
\omega $ for near-extremal conformal-dS black holes in the next section.

\section{QN modes of the near-extremal black holes \label{QNM copy(1)}}

Now, we obtain an exact expression for the near-extremal-dS black holes in
conformal gravity. We start with introducing the normalized variables $x$
and $\tilde{L}^{2}$ as below
\begin{equation}
x=\frac{r}{2M},\;\;\;\;\;\;\tilde{L}^{2}=\frac{L^{2}}{4M^{2}},
\label{normal}
\end{equation}%
where $L$\ is related to the cosmological constant by $L^{2}=3/\Lambda $. In
terms of the new variables, the metric function (\ref{MF1}) can be rewritten
as
\begin{equation}
f\left( x\right) =C_{1}-\frac{1}{x}-\frac{C_{1}^{2}-1}{3}x-\frac{x^{2}}{%
\tilde{L}^{2}}.  \label{xNMF}
\end{equation}

One can find that the spacetime has two horizons: an event horizon ($x=x_{e}$%
) and a cosmological horizon ($x=x_{c_{0}}$) so that $x_{e}<x_{c_{0}}$. The
metric function has three roots at $x_{e}$, $x_{c_{0}}$, and $x_{0}$
(negative root), and therefore, we can express the metric function in terms
of these quantities as follows
\begin{equation}
f\left( x\right) =\frac{1}{x\tilde{L}^{2}}\left( x-x_{e}\right) \left(
x_{c_{0}}-x\right) \left( x-x_{0}\right) .  \label{xMMF}
\end{equation}

Equating Eqs. (\ref{xNMF}) and (\ref{xMMF}), the following relations between
parameters is found
\begin{equation}
x_{0}=-\frac{x_{e}x_{c_{0}}\left( x_{e}+x_{c_{0}}\pm Y\right) }{2x_{e}\left(
x_{c_{0}}-x_{e}\right) +2x_{c_{0}}^{2}\left( x_{e}^{2}-1\right) },
\label{x0}
\end{equation}%
\begin{equation}
\tilde{L}^{2}=-x_{e}x_{c_{0}}x_{0},  \label{xL2}
\end{equation}%
\begin{equation}
C_{1}=\frac{3x_{e}+3x_{c_{0}}\mp Y}{2x_{e}x_{c_{0}}},  \label{xq}
\end{equation}%
\begin{equation}
Y=\sqrt{x_{e}^{2}\left( 4x_{c_{0}}^{2}-3\right) -3x_{c_{0}}\left(
x_{c_{0}}-2x_{e}\right) }
\end{equation}%
where $x_{e}$ and $x_{c_{0}}$ are considered as two fundamental parameters
of the spacetime. Regarding Eqs. (\ref{x0}) and (\ref{xq}), the upper sign
is compatible with the Schwarzschild-dS black hole at $C_{1}\rightarrow 1$
limit \cite{SchwDS}, and thus, we shall consider this sign in future
calculations. In addition, one can obtain the surface gravity at the event
horizon $x_{e}$ by differentiating the metric function (\ref{xMMF})%
\begin{equation}
\tilde{\kappa}_{e}=2M\kappa _{e}=\left. \frac{1}{2}\frac{df(x)}{dx}%
\right\vert _{x=x_{e}}=\frac{(x_{c_{0}}-x_{e})\left( x_{e}-x_{0}\right) }{%
2x_{e}\tilde{L}^{2}}.  \label{xSG}
\end{equation}

Now, we concentrate our attention to the near-extremal black holes and
investigate the possible stability. The near-extremal regime is defined in a
case that the cosmological horizon $x_{c_{0}}$ is very close to the black
hole event horizon $x_{e}$ ($x_{c_{0}}-x_{e}<<x_{e}$). Therefore, we can
consider the following approximations for Eqs. (\ref{x0})-(\ref{xq}) and (%
\ref{xSG})
\begin{equation}
x_{0}\sim -\frac{x_{e}}{x_{e}-1},\;\;\;\tilde{L}^{2}\sim \frac{x_{e}^{3}}{%
x_{e}-1},\;\;\;C_{1}\sim \frac{3-x_{e}}{x_{e}},\;\;\;\tilde{\kappa}_{e}\sim
\frac{(x_{c_{0}}-x_{e})}{2x_{e}^{2}},  \label{xNEP}
\end{equation}%
with $x_{e}>1$ as a constraint. This lower bound on $x_{e}$ is an
interesting feature of the nearly extreme conformal-dS black holes in
conformal gravity. This bound means that the event horizon radius $r_{e}$
(corresponds to $x_{e}$) should be larger than the Schwarzschild radius $%
r_{s}=2M$ ($r_{e}>2M$), which is due to the presence of linear $x$-term in
the metric function (\ref{xNMF}). As a result, there will be some conditions
on the other parameters ($x_{0}<-1$, $\tilde{L}^{2}>27/4$, and $-1<C_{1}<2$%
). Soon, we will find that the condition $x_{e}>1$ puts an upper bound on
the real and imaginary parts of the QN frequencies. If one violates this
constraint, the black hole converts to a naked singularity. Note that if we
choose the lower sign in Eqs. (\ref{x0}) and (\ref{xq}), $x_{0}$ will be
positive and $\tilde{L}^{2}$ will be negative in Eq. (\ref{xNEP}), and
therefore, we have a naked singularity again. Besides, since $x$ varies
between $x_{c_{0}}$ and $x_{e}$, we find
\begin{equation}
x-x_{0}\sim x_{e}-x_{0}\sim \frac{x_{e}^{2}}{x_{e}-1},  \label{xx0}
\end{equation}%
which is the key point to construct exact modes. By using the relation of $%
\tilde{L}^{2}$ from (\ref{xNEP}) and (\ref{xx0}), we can rewrite the near
horizon form of the metric function (\ref{xMMF}) as%
\begin{equation}
f\left( x\right) \sim \frac{\left( x-x_{e}\right) \left( x_{c_{0}}-x\right)
}{x_{e}^{2}}.  \label{xNEMF}
\end{equation}

Here, the normalized tortoise coordinate is
\begin{equation}
\tilde{r}_{\ast }=\frac{r_{\ast }}{2M}=\int \frac{dx}{f(x)},  \label{TDC}
\end{equation}%
and by employing Eqs. (\ref{xNEMF}) and (\ref{TDC}), we can invert the
relation $\tilde{r}_{\ast }(x)$ into the following form%
\begin{equation}
x=\frac{x_{c_{0}}e^{2\tilde{\kappa}_{e}\tilde{r}_{\ast }}+x_{e}}{1+e^{2%
\tilde{\kappa}_{e}\tilde{r}_{\ast }}},  \label{X}
\end{equation}%
for the near-extremal regime. Substituting (\ref{X}) into (\ref{xNEMF}) and
using $\tilde{\kappa}_{e}$ from (\ref{xNEP}), it is straightforward to show
that the metric function (\ref{xNEMF}) converts into
\begin{equation}
f(x)=\frac{\tilde{\kappa}_{e}^{2}x_{e}^{3}}{\left( 3-x_{e}\right) \cosh
^{2}\left( \tilde{\kappa}_{e}\tilde{r}_{\ast }\right) }.  \label{xfx}
\end{equation}

In addition, the wave equation (\ref{WE}) and effective potential (\ref{EP})
are now given by
\begin{equation}
\left[ \partial _{\tilde{r}_{\ast }}^{2}+\tilde{\omega}^{2}-\tilde{V}%
_{l}\left( \tilde{r}_{\ast }\right) \right] \Psi _{l}\left( \tilde{r}_{\ast
}\right) =0,\;\;\;\;\;\tilde{\omega}=2M\omega ,  \label{xWE}
\end{equation}%
and%
\begin{equation}
\tilde{V}_{l}\left( \tilde{r}_{\ast }\right) =4M^{2}V_{l}(r_{\ast })=f\left(
x\right) \left[ \frac{l\left( l+1\right) }{x^{2}}+\frac{1}{x^{3}}-\frac{%
C_{1}^{2}-1}{3x}-\frac{2}{\tilde{L}^{2}}\right] .  \label{xEP}
\end{equation}

By considering (\ref{xfx}) and the relation of $C_{1}$\ and $\tilde{L}^{2}$
given in (\ref{xNEP}), one can express the effective potential (\ref{xEP})
in the near-extremal regime as follows%
\begin{equation}
\tilde{V}_{l}\left( \tilde{r}_{\ast }\right) =\frac{\tilde{\kappa}%
_{e}^{2}l(l+1)}{\cosh ^{2}\left( \tilde{\kappa}_{e}\tilde{r}_{\ast }\right) }%
,
\end{equation}%
and therefore, the wave equation (\ref{xWE}) reduces to
\begin{equation}
\left[ \partial _{\tilde{r}_{\ast }}^{2}+\tilde{\omega}^{2}-\frac{\tilde{V}%
_{0}}{\cosh ^{2}\left( \tilde{\kappa}_{e}\tilde{r}_{\ast }\right) }\right]
\Psi _{l}\left( \tilde{r}_{\ast }\right) =0,  \label{xNEWE}
\end{equation}%
with
\begin{equation}
\tilde{V}_{0}=\tilde{\kappa}_{e}^{2}l(l+1).
\end{equation}

The effective potential in (\ref{xNEWE}) is known as the P\"{o}shl-Teller
potential \cite{Poshl}. Solutions of wave equation (\ref{xNEWE}) have been
investigated by considering the proper boundary conditions \cite{Ferrari}
\begin{equation}
\left\{
\begin{array}{c}
\Psi _{l}\left( \tilde{r}_{\ast }\right) \sim e^{-i\tilde{\omega}\tilde{r}%
_{\ast }}\ \ \ \ as\ \ \ \ \ \ \ \ \tilde{r}_{\ast }\rightarrow -\infty \
(x\rightarrow x_{e}), \\
\Psi _{l}\left( \tilde{r}_{\ast }\right) \sim e^{i\tilde{\omega}\tilde{r}%
_{\ast }}\ \ \ \ \ \ \ as\ \ \ \ \ \ \ \ \ \tilde{r}_{\ast }\rightarrow
\infty \ (x\rightarrow x_{c_{0}}),%
\end{array}%
\right.
\end{equation}%
mean the wave at the event horizon is purely incoming and at the
cosmological horizon is purely outgoing. The frequencies $\tilde{\omega}$'s
that satisfy the boundary conditions are the QNFs. For the P\"{o}shl-Teller
potential, one can show that the QNFs are given by (see \cite{Ferrari} for
more details)
\begin{equation}
\tilde{\omega}=\tilde{\kappa}_{e}\left[ \sqrt{l\left( l+1\right) -\frac{1}{4}%
}-\left( n+\frac{1}{2}\right) i\right] ;\ \ n=0,1,2,...  \label{xFF}
\end{equation}

Regarding Eq. (\ref{xFF}), one can find that the multipole (overtone) number
does not affect the imaginary (real) part of the frequencies. As a result,
one can obtain the real part of QNFs with a large imaginary part which is
equal to the Barbero-Immirzi parameter \cite{HodPRL,Dreyer,Corichi,Motl}.
The Barbero-Immirzi parameter \cite{Immirzi} is a factor introduced by hand
in order that Loop Quantum Gravity reproduces the black hole entropy.

In order to confirm the correctness of the obtained relation in Eq. (\ref%
{xFF}), we calculated the QNMs for some special cases by using the sixth
order WKB formula \cite{Schutz,Iyer,KonoplyaWKB} and the asymptotic
iteration method (AIM) \cite{Naylor}. The results are presented in table $I$%
. This table shows a remarkable agreement between the WKB formula, AIM and
the results of Eq. (\ref{xFF}). The calculated frequencies from WKB and AIM
get closer to the results of Eq. (\ref{xFF}) whenever $x_{c_{0}}$ approaches
to $x_{e}$, and we expected this behavior for an exact formula at the
near-extremal regime. Therefore, we ensure that the obtained relation (\ref%
{xFF}) is indeed correct up to terms of order $\mathcal{O}(x_{c_{0}}-x_{e})$
or higher and it is an exact expression for the QNMs of nearly extreme
conformal-dS black holes undergoing massless scalar perturbations.

\begin{center}
\begin{tabular}{|c|c|c|c|c|c|c|}
\hline\hline
$\frac{\tilde{\omega}}{\tilde{\kappa}_{e}}\left( \text{Eq.(\ref{xFF})}%
\right) $ & $\frac{\tilde{\omega}}{\tilde{\kappa}_{e}}\ ($WKB$)$ & $\frac{%
\tilde{\omega}}{\tilde{\kappa}_{e}}\ ($AIM$)$ & $l$ & $C_{1}$ & $\tilde{L}$
& $x_{c_{0}}-x_{e}$ \\ \hline\hline
$1.323-0.5000i$ & $1.322-0.4996i$ & $1.322-0.4995i$ & $1$ & $1.781953$ & $4$
& $0.0021452$ \\ \hline
$2.398-0.5000i$ & $2.396-0.4995i$ & $2.396-0.4995i$ & $2$ & $1.781953$ & $4$
& $0.0021452$ \\ \hline
$1.323-0.5000i$ & $1.316-0.4976i$ & $1.316-0.4975i$ & $1$ & $1.781850$ & $4$
& $0.0118026$ \\ \hline
$2.398-0.5000i$ & $2.386-0.4975i$ & $2.386-0.4975i$ & $2$ & $1.781850$ & $4$
& $0.0118026$ \\ \hline\hline
\end{tabular}

Table $I$: The fundamental QNMs calculated by Eq. (\ref{xFF}), sixth order
WKB formula, and AIM after $15$ iterations.
\end{center}

\begin{figure}[tbp]
$%
\begin{array}{ccc}
\epsfxsize=7.5cm \epsffile{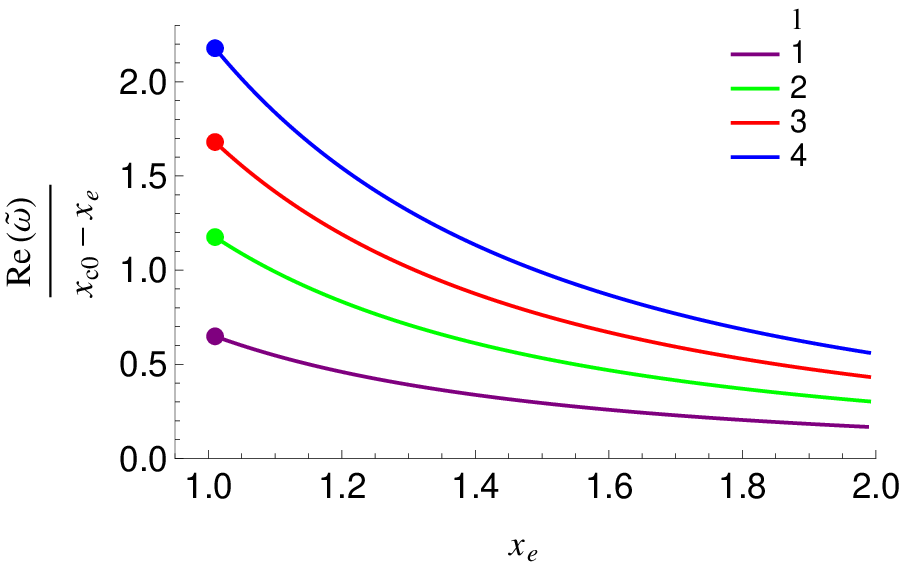} & \epsfxsize=7.5cm \epsffile{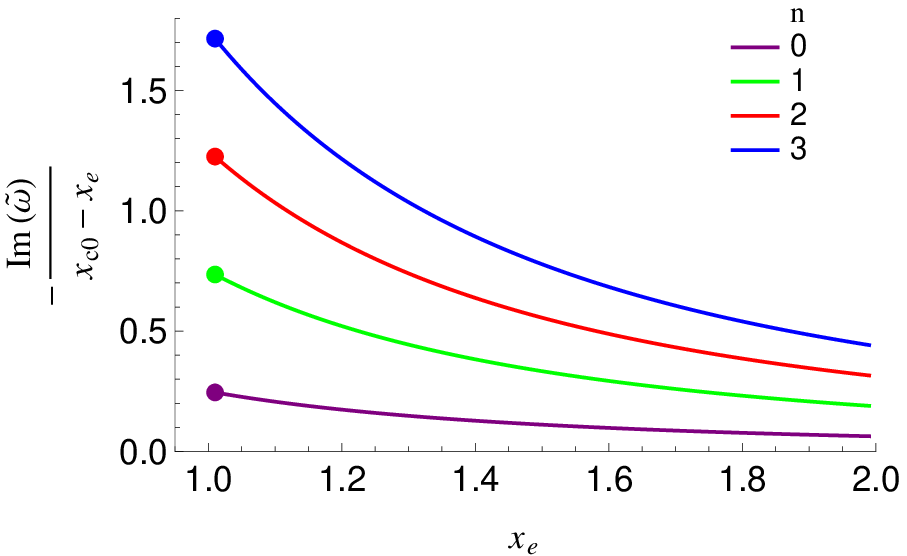}
&
\end{array}
$%
\caption{The real and imaginary parts of the QNMs as functions of the
normalized event horizon radius $x_{e}$. The circular points shows the
maximum value of frequencies.}
\label{xeFig}
\end{figure}

Figure \ref{xeFig} shows the behavior of the real and imaginary parts of
frequencies versus the normalized event horizon radius $x_{e}$. Since $%
x_{e}>1$, the lower value of $x_{e}$\ puts an upper bound on the values of
frequencies for fixed $l$\ and $n$. In addition, the effective potential in
Eq. (\ref{xNEWE}) is positive everywhere, and therefore, the black holes are
dynamically stable.

\subsection{Geodesic instability and QN modes}

At this stage, we obtain the parameters of the circular null geodesics (the
angular velocity $\Omega _{c}$ at the unstable circular null geodesic and
the Lyapunov exponent $\lambda $) to check the validity of correspondence
between these parameters and QNFs in the eikonal limit. In the eikonal limit
($l\rightarrow \infty $), the frequencies (\ref{xFF}) reduce to
\begin{equation}
\tilde{\omega}=\tilde{\kappa}_{e}\left[ l-\left( n+\frac{1}{2}\right) i%
\right] ;\ \ n=0,1,2,...,  \label{xeikonal}
\end{equation}

A circular null geodesic for an arbitrary metric function $f(x)$ satisfies
the following condition \cite{Geodesic}
\begin{equation}
f_{c}=Mx_{c}f_{c}^{\prime },  \label{condition}
\end{equation}%
in which the subscript $c$ refers to the radius $x=x_{c}$ of a circular null
geodesic. One can obtain $x_{c}$ by using Eqs. (\ref{xNMF}) and (\ref%
{condition}) as
\begin{equation}
x_{c}=\frac{3}{C_{1}\pm 1},  \label{xc}
\end{equation}%
which we choose the upper sign to be compatible with $C_{1}\rightarrow 1$
limit (Schwarzschild-dS case \cite{Geodesic}). Moreover, the angular
velocity $\Omega _{c}$ for an arbitrary metric function $f(x)$ is given by
\cite{Geodesic}
\begin{equation}
\tilde{\Omega}_{c}^{2}=4M^{2}\Omega _{c}^{2}=\frac{f_{c}}{x_{c}^{2}},
\label{av}
\end{equation}%
in which for our black hole case study converts to
\begin{equation}
\tilde{\Omega}_{c}^{2}=-\frac{1}{\tilde{L}^{2}}+\frac{1}{x_{c}^{2}}\left(
\frac{2}{x_{c}}-C_{1}\right) ,  \label{av2}
\end{equation}%
and we regarded Eq. (\ref{condition}) in order to obtain this equation.
Taking (\ref{xL2}), (\ref{xq}), (\ref{xNEP}), and (\ref{xc}) into account,
one can find $\tilde{\Omega}_{c}$ with the following explicit form%
\begin{equation}
\tilde{\Omega}_{c}=\tilde{\kappa}_{e}.  \label{xAV}
\end{equation}

Now, we obtain the Lyapunov exponent $\lambda $ for the near-extremal
conformal-dS black holes. Using Eqs. ($1$), ($27$), ($34$), and ($36$) of
Ref. \cite{Geodesic}, it is straightforward to show that the Lyapunov
exponent for an arbitrary metric function $f(x)$ takes the following form
\begin{equation}
\tilde{\lambda}=\sqrt{\frac{f_{c}}{x_{c}^{2}}\left(
f_{c}-2M^{2}x_{c}^{2}f_{c}^{\prime \prime }\right) },\ \ \ \ \ \tilde{\lambda%
}=2M\lambda ,  \label{le}
\end{equation}%
which for these black holes reduces to $\tilde{\lambda}^{2}=\tilde{\Omega}%
_{c}^{2}\left( C_{1}-\frac{C_{1}^{2}-1}{3}x_{c}\right) $ versus the angular
velocity. Considering (\ref{xL2}), (\ref{xq}), (\ref{xNEP}), (\ref{xc}), and
(\ref{av2}), the final result is%
\begin{equation}
\tilde{\lambda}=\tilde{\kappa}_{e},  \label{xLE}
\end{equation}

Comparing Eq. (\ref{xeikonal}) with Eqs. (\ref{xAV}) and (\ref{xLE}), we
deduce
\begin{equation}
\tilde{\omega}=\tilde{\Omega}_{c}l-\tilde{\lambda}\left( n+\frac{1}{2}%
\right) i;\ \ n=0,1,2,...,
\end{equation}%
which shows that the real and imaginary parts of the QNFs in the eikonal
limit, respectively, are given by the frequency and the inverse of
instability timescale of the unstable circular null geodesics, and
therefore, the correspondence is guaranteed for the massless scalar
perturbation of CG.

\section{Conclusions \label{Conclusions}}

We have considered a minimally coupled massless scalar perturbation in the
background spacetime of the conformal-dS black holes and found an analytical
expression for the near horizon QNMs of these solutions which was correct up
to terms of order $\mathcal{O}(x_{c_{0}}-x_{e})$ or higher. In addition, we
calculated the QNFs for some special cases by using the sixth order WKB
approximation and the AIM (after $15$ iterations). We have seen that the
analytical expression was in a remarkable agreement with their results which
confirmed that the obtained exact formula is indeed correct. In addition,
the effective potential was positive everywhere which shows that these black
holes are dynamically stable.

Moreover, we found that the multipole (overtone) number did not affect the
imaginary (real) part of the frequencies. Therefore, one can obtain an $l$%
-dependent real part of the QNFs with an arbitrary large imaginary part
which is equal to the Barbero-Immirzi parameter. We have also seen that the
near-extremal-dS black hole in conformal gravity suffered an lower bound on
its event horizon radius which led to an upper bound on the values of the
real and imaginary parts of frequencies for fixed $n$ and $l$.

On the other hand, we have calculated the parameters of the unstable
circular null geodesics, i.e., the angular velocity $\Omega _{c}$ at the
unstable circular null geodesic and the Lyapunov exponent $\lambda $ which
is related to the instability timescale of the geodesic motion. Then, it was
shown that, unlike the Gauss-Bonnet gravity, the correspondence between
these parameters and QNFs in the eikonal limit is guaranteed for massless
scalar perturbations of the conformal-dS black holes in the nearly extreme
regime.

Now, we finish our paper with some suggestions. Investigating the
near-extremal black holes in dRGT massive gravity can also be an interesting
work due to probability of existence of bound on the lower and/or upper
values of the frequencies (because of the structure of its metric function).
In addition, it has been shown that the near-extremal Reissner-Nordstr\"{o}%
m-dS black holes have an instability spectra in the frequencies \cite{Hod}
when you consider charged scalar perturbations. Therefore, one can study the
charged black holes of conformal gravity in the nearly extreme regime and
search for instability conditions by investigating a charged scalar
perturbation. Finally, checking the validity of the correspondence between
quasinormal modes and unstable circular null geodesics for mentioned works
is also recommended.

\section{acknowledgements}

The authors wish to thank the anonymous referee for the constructive
comments that enhanced the quality of this paper. We wish to thank Shiraz
University Research Council. This work has been supported financially by the
Research Institute for Astronomy and Astrophysics of Maragha, Iran.

\end{document}